\documentclass[twocolumn,showpacs,preprintnumbers,amsmath,amssymb]{revtex4}
\usepackage{graphicx}

\begin{document}
\draft
\title{Coherent tunneling by adiabatic passage
in an optical waveguide system}
 \normalsize
\author{S. Longhi, G. Della Valle, M. Ornigotti, and P. Laporta}
\address{Dipartimento di Fisica and Istituto di Fotonica e Nanotecnologie del CNR,
Politecnico di Milano, Piazza L. da Vinci 32,  I-20133 Milan,
Italy}

%
\bigskip
\begin{abstract}
\noindent We report on the first experimental demonstration of
light transfer in an engineered triple-well optical waveguide
structure which provides a classic analogue of Coherent Tunnelling
by Adiabatic Passage (CTAP) recently proposed for coherent
transport {\it in space} of neutral atoms or electrons among
tunneling-coupled optical traps or quantum wells [A.D. Greentree
{\it et al.}, Phys. Rev. B {\bf 70}, 235317 (2004); K. Eckert {\it
et al.} Phys. Rev. A {\bf 70}, 023606 (2004)]. The direct
visualization of CTAP wavepacket dynamics enabled by our simple
optical system clearly shows that in the counterintuitive passage
scheme light waves tunnel between the two outer wells without
appreciable excitation of the middle well.
\end{abstract}



\maketitle

\newpage
Similarities between quantum and classical phenomena are not
uncommon in physics albeit quantum and classical physics are
grounded on different paradigms. Such analogies, which have been
regarded as a mere curiosity until quite recently, have been
fruitfully exploited in the past recent years in emerging research
areas such as quantum computing, nano-devices and new forms of
light (see \cite{Dragoman} and references therein). For instance,
in its early developments the field of photonic crystals has
borrowed many ideas from solid-state physics
\cite{Joannopoulos97}. Quantum-classical analogies in apparently
unrelated fields have been also successfully exploited to mimic at
a macroscopic level many quantum phenomena which are currently not
of easy access in microscopic quantum systems. In particular,
engineered photonic structures have proven to be a useful
laboratory tool to investigate the optical analogues of a wide
variety of quantum effects, including among others the optical
analogues of Bloch oscillations
\cite{Morandotti99,Sapienza03,Christodoulides03,Trompeter06a,Trompeter06b},
Zener tunneling \cite{Trompeter06a,Pavesi05}, dynamic localization
\cite{Longhi06a}, coherent enhancement and destruction of
tunneling \cite{Longhi07}, adiabatic stabilization of atoms in
strong fields \cite{Longhi05}, and Anderson localization \cite{Segev07}.\\
On the other hand, the ability of coherently manipulating the
state evolution of a quantum system and of transferring a particle
between positional quantum states in a controllable and reliable
way is currently a subject of great relevance in quantum physics
\cite{Shapiro07}. Recently, a robust scheme for coherent quantum
transport {\it in space} of the wave function among
tunneling-coupled quantum wells has been independently proposed
for neutral atoms in optical traps \cite{Eckert04} and for
electrons in quantum dot systems \cite{Greentree04} (see also
\cite{addatoms,adddots}). In such a scheme, which was referred to
as "Coherent Tunneling Adiabatic Passage" (CTAP)
\cite{Greentree04}, the tunneling interaction between adjacent
quantum units is dynamically tuned by changing either the distance
or the height of the neighboring potential wells following a
counterintuitive scheme which is reminiscent of the celebrated
stimulated Raman adiabatic passage (STIRAP) technique
\cite{Bergmann98,Vitanov01}, originally developed for transferring
population between two long-lived atomic or molecular energy
levels optically connected to a third auxiliary state. To date,
however, no experimental demonstration of coherent transport in
space of a quantum particle by adiabatic passage
has been reported.\\
In this Letter we show an experimental visualization of coherent
transport of light waves by adiabatic passage in an engineered
optical waveguide system which provides the optical analogue
\cite{Kenis01,Longhi06,Paspalakis06} of quantum CTAP in a dynamic
triple-well potential. Our classic analogue of CTAP enables a
simple and direct visualization of the coherent wave
packet transport process, which is a rather unique possibility offered by optics \cite{Trompeter06a}.\\
\begin{figure}
\includegraphics[scale=0.45]{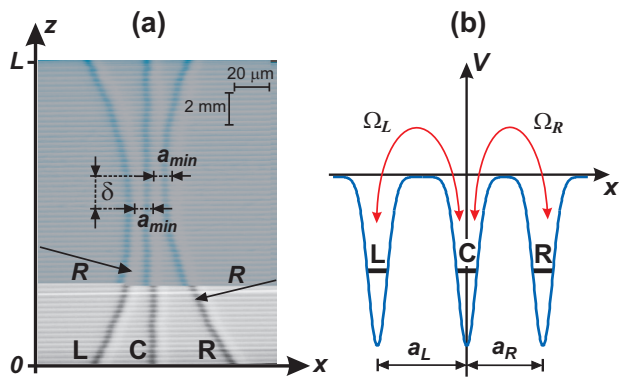}
\caption{(color online) (a) Microscope image (top view) of the
manufactured three-waveguide structure. (b) Schematic of light
transfer in a triple well system with controlled hopping rates.}
\end{figure}
The optical structure we designed and manufactured consists of a
set of three $L=24$-mm-long channel waveguides in the geometry
shown in Fig.1(a). Light trapped in one of the three waveguides
may tunnel into the neighbor waveguide, the tunneling rate
$\Omega$ being controlled in our experiment by the waveguide
distances. The left (L) and right (R) waveguides are enough far
apart each other that direct tunneling between them is fully
negligible over the sample length. As the central (C) waveguide is
straight, the left and right waveguides are weakly curved, with
opposite curvature of radius $R$, and longitudinally displaced
each other by a value $\delta>0$ so that their minimum distance
$a_{min}$ from the central waveguide is reached at different
propagation lengths [see Fig.1(a)]. The distances $a_L(z)$ and
$a_R(z)$ of the left and right waveguides from the central one
change along the paraxial propagation distance $z$, from the input
($z=0$) to the output ($z=L$) planes, according to the relations
$a_L(z) \simeq a_{min}+(z-L/2)^2/(2R)$ and $a_R(z) \simeq
a_{min}+(z-L/2-\delta)^2/(2R)$. With a proper engineering of the
coupling rate between adjacent waveguides, the designed coupling
geometry is suited to mimic CTAP. The quantum-optical analogy is
at best captured starting from the Schr\"{o}dinger-like paraxial
wave equation describing monochromatic light propagation at
wavelength $\lambda$ along the paraxial direction $z$ of the
optical structure (see, for instance, \cite{Longhi07})
\begin{equation} i \lambdabar \frac{\partial
\psi}{\partial z} = -\frac{\lambdabar^2}{2n_s} \nabla^{2}_{x,y}
\psi + V(x,y,z) \psi.
\end{equation}
where $\lambdabar \equiv \lambda / (2 \pi)$ is the reduced
wavelength, $V(x,y,z)=[n_{s}^2-n^2(x,y,z)]/(2 n_s) \simeq
n_s-n(x,y,z)$ is the $z-$dependent triple-well potential,
$n(x,y,z)$ is the refractive index profile of the three-waveguide
system at plane $z$, $n_s$ is the reference (substrate) refractive
index, and $|\psi|^2$ is proportional to the local light
intensity. Indicating by $c_L$, $c_C$ and $c_R$ the amplitudes of
the fundamental modes in the three waveguides, a coupled-mode
equation analysis of Eq.(1) leads to the following STIRAP-like
equations describing optical tunneling among the three waveguides
\cite{Paspalakis06,Longhi06}
\begin{figure}
\includegraphics[scale=1]{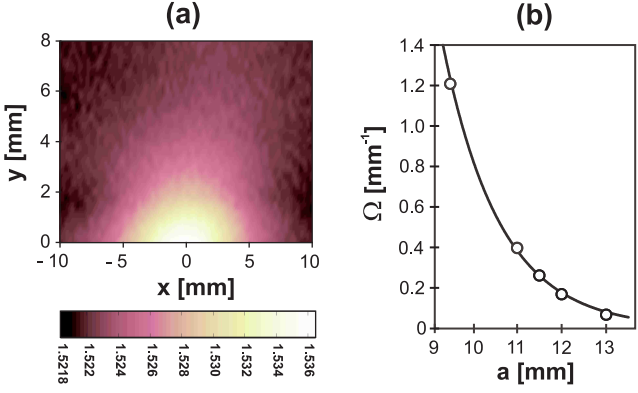}
\caption{(a) (color online) Measured 2D refractive index profile
$n_w(x,y)$ of the Ag-Na diffused waveguide. (b) Measured tunneling
rate $\Omega$ between two waveguides versus waveguide distance $a$
(circles). The solid curve is an exponential fit of the
experimental data.}
\end{figure}
\begin{eqnarray}
i \frac{dc_{L}}{dz}  & = &    -\Omega_{L}(z) c_C \\
i \frac{dc_C}{dz} & = &    -\Omega_{L}(z) c_L-  \Omega_{R}(z) c_R
\\
i \frac{dc_{R}}{dz}  & = &    -\Omega_{R}(z) c_C
\end{eqnarray}
where $\Omega_L(z)=\Omega(a_L(z))$,
$\Omega_R(z)=\Omega(a_R(z))=\Omega_L(z-\delta)$ and $\Omega(a)$ is
the tunneling rate between two adjacent waveguides placed at
distance $a$. Note that the temporal variable of the analog atomic
STIRAP process \cite{Bergmann98} is here played by the spatial
propagation distance $z$, whereas the Rabi frequencies of pump and
Stokes pulses correspond to the $z$-dependent tunneling rates
$\Omega_L(z)$ and $\Omega_R(z)$ [see Fig.1(b)]. Note also that, as
for $\delta>0$ $\Omega_L(z)$ precedes
$\Omega_R(z)=\Omega_L(z-\delta)$, the counterintuitive (intuitive)
pulse sequence of atomic STIRAP is simulated here by exciting the
right (left) waveguide at $z=0$. Light transfer from the right to
the left waveguides leaping over the central one is related to the
existence of a dark state for Eqs.(2-4),  given by
$(c_L,c_C,c_R)=(\Omega_R/\sqrt{\Omega_R^2+\Omega_L^2},0,-\Omega_L/\sqrt{\Omega_R^2+\Omega_L^2})$.
Light transfer is simply achieved by adiabatic change of the
tunneling rates so that at the initial and final propagation
planes the dark state corresponds to light trapped in the right
and in the left waveguides, respectively.\\
 \begin{figure}
\includegraphics[scale=1.2]{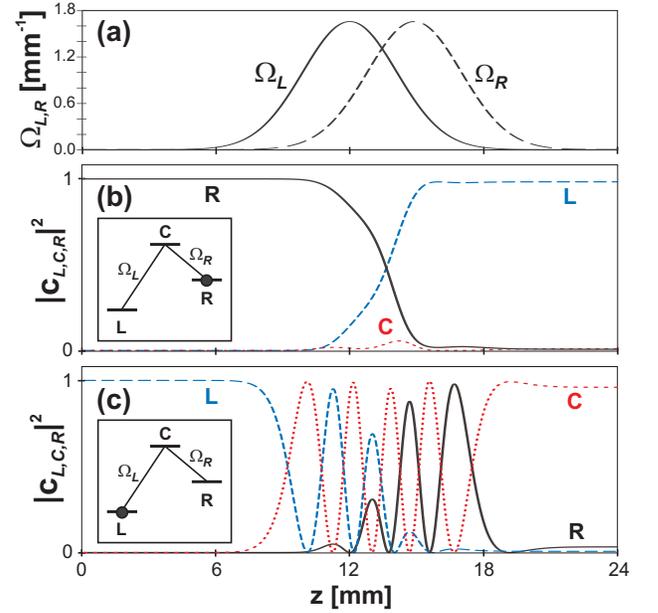}
\caption{(color online) (a) Behavior of the tunneling rates
$\Omega_L$ and $\Omega_R$ versus propagation distance $z$ and
behavior of fractional light power trapped in the three waveguides
L, C and R versus $z$ as obtained by coupled-mode equation
analysis corresponding to: (b) a counterintuitive pulse sequence
(STIRAP), and (c) an intuitive pulse sequence. Parameter values
are given in the text.}
\end{figure}
 \begin{figure}
\includegraphics[scale=1.2]{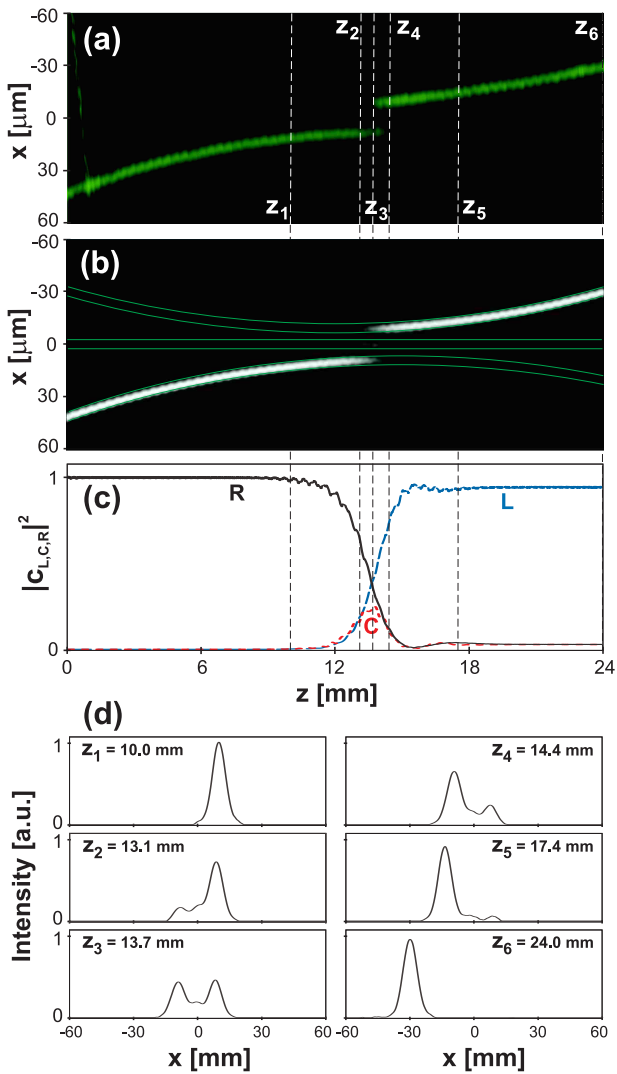}
\caption{(color online) (a) Measured fluorescence pattern as
recorded on the CCD camera (top view of the sample) corresponding
to the input excitation of the right waveguide (CTAP). (b)
Behavior of light intensity distribution $|\psi(x,0,z)|^2$
predicted by a numerical analysis of Eq.(1), and (c) corresponding
fractional beam power trapped in the three waveguides versus
propagation distance $z$. In (d) a few transverse cross sections
of the measured fluorescence pattern are depicted at propagation
distances corresponding to the vertical dashed lines in Fig.4(a)}
\end{figure}
In our experiment, the waveguides were fabricated by the Ag-Na ion
exchange technique  and applying a titanium mask onto an
Er:Yb-doped glass substrate (Schott IOG1) \cite{DellaValle06}. The
2D index profile $n_w(x,y)$ of the single waveguide, shown in
Fig.2(a), was measured using a refracted-near-field profilometer
(Rinck Elektronik) at a wavelength of 670 nm and fitted by the
relation
\begin{equation}
 n_w(x,y) \simeq n_s+\Delta n [g(x-a/2)+g(x+a/2)]f(y),
\end{equation}
where $\Delta n \simeq 0.0148$ is peak index change, $g(x)=[{\rm
erf}((x+w)/D_x)-{\rm} {\rm erf}((x-w)/D_x)]/[2 {\rm erf}(w/D_x)]$
and $f(y)=[1-{\rm erf}(-y/D_y)]$ define the shape of the index
profile parallel to the surface of the waveguide ($x$-direction)
and perpendicular to the surface ($y$-direction), respectively,
$2w \simeq 5 \; \mu$m is the channel width and $D_x \simeq 4.6 \;
\mu$m, $D_y \simeq 4.1 \; \mu$m  are the lateral and in-depth
diffusion lengths. The waveguides have been excited at $\lambda
\simeq 980$ nm wavelength using a single-frequency tunable laser
diode fiber-coupled to the polished input facet of the sample. The
probing light is partially absorbed by the Yb$^{3+}$ ions
(absorption length $\sim 6$ mm), yielding a bright green
upconversion luminescence arising from the radiative decay of
higher-lying energy levels of Er$^{3+}$ ions which is proportional
to the local photon density of the probing radiation. The
fluorescence was recorded from the top of the sample using a CCD
camera connected to an optical microscope, providing a
magnification factor of $\sim 12$. The microscope was mounted onto
a PC-controlled micropositioning system, and successive
acquisitions of the fluorescence images along the paraxial
propagation distance $z$, at steps of $400 \; \mu$m, allowed us to
visualize step-by-step the flow of light along the three-waveguide
system under different excitation conditions. The single-mode
coupling fiber is mounted on a computer-controlled positioning
system, which allows for a precise control of the input beam
launching conditions. Excitation of the fundamental mode of either
the left or the right waveguide with good mode matching is simply
accomplished by scanning the fiber transversely
along the input facet of the sample.\\
The parameters $R$, $a_{min}$ and $\delta$ which determine the
behavior of tunneling rates $\Omega_{L,R}(z)$ have been designed
in order to fulfill the adiabaticity criterion requested for the
observation of STIRAP \cite{Bergmann98}. To this aim, we
preliminarily performed a set of  accurate experimental
measurements to determine the tunneling rate $\Omega(a)$ of two
adjacent waveguides versus their distance $a$. This was
accomplished by manufacturing five two-waveguide straight
couplers, with different waveguide interspacing ($a=9.5~\mu$m,
$a=11.0~\mu$m, $a=11.5~\mu$m, $a=12.0~\mu$m and $a=13.0~\mu$m),
and by measuring the spatial period of the light tunneling pattern
observed on the top of the sample. The experimental results, shown
in Fig.2(b), indicate that the tunneling rate dependence versus
distance $a$ may be well fitted by the exponential curve
 $\Omega(a)=\Omega_0 \exp[-\gamma (a-a_0)]$ with
$\Omega_0 \simeq 1.789~{\rm mm}^{-1}$ and $\gamma \simeq 0.7759~
\mu{\rm m}^{-1}$, at least for distances $a \gtrsim a_0 =
9$~$\mu$m. Using such a relation, a possible combination of
parameter values for $a_{min}$, $R$ and $\delta$ that ensure
adiabaticity of the STIRAP process and optimal overlap between
pump and probe pulses was determined from a numerical analysis of
the coupled-mode equations (2-4). As an example, Fig.3 shows the
numerically-computed behavior of the normalized mode power
$|c_{L}|^2$, $|c_{C}|^2$ and $|c_{R}|^2$ trapped in the three
waveguides versus propagation distance $z$ for an initial
excitation of the right waveguide, corresponding to a
counterintuitive pulse sequence (STIRAP) [Fig.3(b)], and for the
excitation of the left waveguide, corresponding to an intuitive
pulse sequence [Fig.3(c)]. Parameter values used in the
simulations, which correspond to the manufactured structure shown
in Fig.1(a), are $R=3.445$~m, $a_{min}=9.1~\mu$m, and
$\delta=2.9$~mm. Figures 4(a) and 5(a) show the measured
fluorescence patterns as obtained by excitation of the right
(Fig.4) and left (Fig.5) waveguides. In Fig.4, the transverse
cross section profiles of the measured fluorescence pattern at a
few propagation distances are also depicted. In Figs.4(b) and
5(b), we plotted the corresponding images of light pattern
evolution $|\psi(x,0,z)|^2$ as predicted by direct numerical
simulations of the paraxial wave equation (1) using a standard
beam propagation software (BeamPROP 5.0) and assuming $n(x,y,z)
\simeq n_w(x,y)+n_w(x-a_L(z),y)+n_w(x+a_R(z),y)$ for the index
profile. The fractional light power trapped in the three
waveguides versus propagation distance $z$, as obtained by the
beam propagation analysis, is also reported. Note that the
experimental results are very well reproduced by the numerical
simulations. In particular, in the counterintuitive tunneling
scheme of Fig.4 the phenomenon of CTAP is clearly demonstrated
since a high-efficiency transfer or light from the right to the
left waveguides is observed with small excitation of the central
waveguide. Conversely, in the intuitive tunneling sequence of
Fig.5 a complex tunneling scenario among the three waveguides,
with most of the power finally left in the central waveguide, is
observed and well reproduced by the numerical analysis. It should
be finally noted that the fraction of light power transiently
trapped in the central waveguide and visible in Fig.4, though
relatively small, could be further reduced using longer
waveguides, the limit to the transient power in the middle
waveguide being ultimately set by the adiabaticity of the process
(see \cite{Bergmann98,Vitanov01}).\\
 \begin{figure}
\includegraphics[scale=1.2]{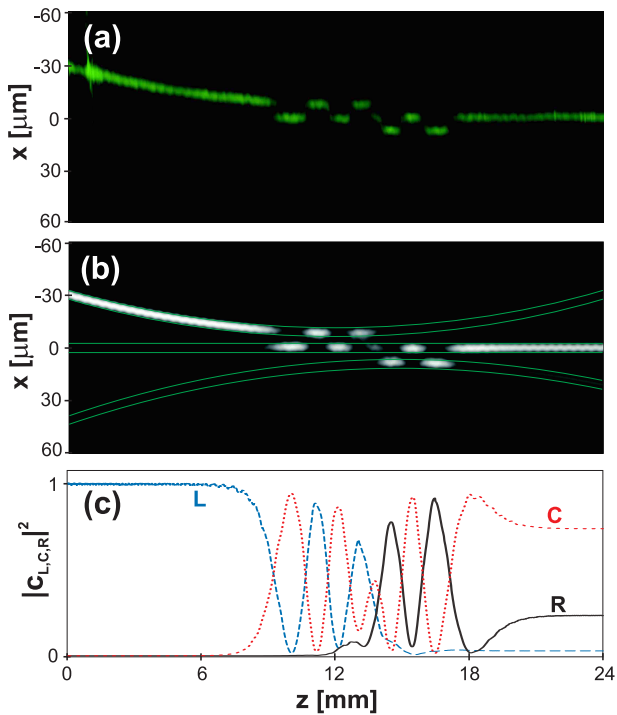}
\caption{(color online) Same as Fig.4, but for excitation of the
left waveguide at the input plane (intuitive pulse sequence).}
\end{figure}
In conclusion, we reported on a classic wave optics analogue of
coherent transport of quantum particles by adiabatic passage,
recently proposed as a robust mean to transfer neutral atoms or
electrons in space \cite{Eckert04,Greentree04}. Our optical
analogy enabled a direct visualization of the tunneling transfer
process and clearly demonstrated that in the scheme mimicking the
counterintuitive pulse sequence of atomic STIRAP light waves are
transferred to the outer wells with small excitation of the
central one.


\begin{thebibliography}{99}

\bibitem{Dragoman}
D. Dragoman and M. Dragoman, {\it Quantum-Classical Analogies}
(Springer, Berlin, 2004).

\bibitem{Joannopoulos97}
J. D. Joannopoulos, P.R. Villeneuve, and S. Fan, Nature {\bf 386},
143 (1997).

\bibitem{Morandotti99}
R. Morandotti, U. Peschel, J.S. Aitchison, H.S. Eisenberg, Y.
Silberberg, Phys. Rev. Lett. {\bf 83}, 4756 (1999); T. Pertsch, P.
Dannberg, W. Elflein, A. Br\"{a}uer, F. Lederer, Phys. Rev. Lett.
{\bf 83}, 4752 (1999).

\bibitem{Sapienza03}
R. Sapienza {\it et al.}, Phys. Rev. Lett. {\bf 91}, 263902
(2003).

\bibitem{Christodoulides03}
D. N. Christodoulides, F. Lederer, and Y. Silberberg, Nature {\bf
424}, 817 (2003).

\bibitem{Trompeter06a}
H. Trompeter {\it et al.}, Phys. Rev. Lett. {\bf 96}, 023901
(2006).

\bibitem{Trompeter06b}
H. Trompeter {\it et al.}, Phys. Rev. Lett. {\bf 96}, 053903
(2006).

\bibitem{Pavesi05}
M. Ghulinyan, C.J. Oton, Z. Gaburro, L. Pavesi, C. Toninelli, and
D. S. Wiersma, Phys. Rev. Lett. {\bf 94}, 127401 (2005).

\bibitem{Longhi06a}
S. Longhi {\it et al.},  Phys. Rev. Lett. {\bf 96}, 243901 (2006);
R. Iyer, J. S. Aitchison, J. Wan, M. M. Dignam, and C.M. de
Sterke, Opt. Express {\bf 15}, 3212 (2007).

\bibitem{Longhi07}
I. Vorobeichik, E. Narevicius, G. Rosenblum, M. Orenstein, and N.
Moiseyev, Phys. Rev. Lett. {\bf 90}, 176806 (2003); G. Della
Valle, M. Ornigotti, E. Cianci, V. Foglietti, P. Laporta, and S.
Longhi, Phys. Rev. Lett. {\bf 98}, 263601 (2007).

\bibitem{Longhi05}
S. Longhi {\it et al.}, Phys. Rev. Lett. {\bf 94}, 073002 (2005).

\bibitem{Segev07}
T. Schwartz, G. Bartal, S. Fishman, and M. Segev, Nature {\bf
446}, 55 (2007).

\bibitem{Shapiro07}
P. Kral, I. Thanopulos, and M. Shapiro, Rev. Mod. Phys. {\bf 79},
53 (2007).

\bibitem{Eckert04}
K. Eckert, M. Lewenstein, R. Corbalan, G. Birkl, W. Ertmer, and J.
Mompart, Phys. Rev. A {\bf 70}, 023606 (2004).

\bibitem{Greentree04}
A.D. Greentree, J.H. Cole, A.R. Hamilton, and L.C.L. Hollenberg,
Phys. Rev. B {\bf 70}, 235317 (2004).

\bibitem{addatoms}
K. Eckert, J. Mompart, R. Corbalan, M. Lewenstein, and G. Birkl,
Opt. Commun. {\bf 264}, 264 (2006); E.M. Graefe, H.J. Korsch, and
D. Witthaut, Phys. Rev. A {\bf 73}, 013617 (2006).

\bibitem{adddots}
J. Siewert and T. Brandes, Advan. Solid-State Phys. {\bf 44}, 181
(2004); L. C. L. Hollenberg, A. D. Greentree, A. G. Fowler, and C.
J. Wellard, Phys. Rev. B {\bf 74}, 045311 (2006); U. Hohenester,
J. Fabian, and F. Troiani, Opt. Commun. {\bf 264}, 426 (2006).

\bibitem{Bergmann98}
K. Bergmann, H. Theuer, and B.W. Shore, Rev. Mod. Phys. {\bf 70},
1003 (1998).

\bibitem{Vitanov01}
N.Y. Vitanov, T. Halfmann, B.W. Shore, and K. Bergmann, Annu. Rev.
Phys. Chem. {\bf 52}, 753 (2001).

\bibitem{Kenis01}
A.M. Kenis, I. Vorobeichik, M. Orenstein, and N. Moiseyev, IEEE J.
Quant. Electron. {\bf 37}, 1321 (2001).

\bibitem{Longhi06}
S. Longhi, Phys. Rev. E {\bf 73}, 026607 (2006); S. Longhi, Phys.
Lett. A {\bf 359}, 166 (2006).

\bibitem{Paspalakis06}
E. Paspalakis, Opt. Commun. {\bf 258}, 31 (2006).

\bibitem{DellaValle06}
G. Della Valle, S. Taccheo, P. Laporta, G. Sorbello, E. Cianci,
and V. Foglietti, Electron. Lett. {\bf 42}, 632 (2006).


\end{thebibliography}
\end{document}